# An Estimate of the Surface Pollution of the Arctic Sea Ice

A. Laubereau and H. Iglev

Physik-Department, Technische Universitaet Muenchen, Muenchen, Germany

Abstract

The Arctic sea ice represents an important energy reservoir for the climate of the northern hemisphere. The shrinking of the polar ice in the past decades decreases the stored energy and raises serious concerns about future climate changes.[1-4] Model calculations of the present authors [5,6] suggest that half of the global warming during the past fifty years is directly related to the retreat of the sea ice, while the cause is not well understood, e.g. the role of surface pollution [7-10]. We have analysed the reported annual melting and freezing data of the northern sea ice in the years 1979 to 2018 [11] to gain some insight. Two features can be deduced from our simple model: (i) recent results [12,13] are confirmed that approximately 60 % of the loss of sea ice stems from energy transport to the arctic region. (ii) We find evidence that the remaining part of the ice retreat originates from an increasing surface absorption of solar radiation, obviously due to the rising surface pollution of the sea ice. While the phenomenon was previously considered by several authors in a qualitative way, our analysis contributes semi-quantitative information on the situation. We estimate that the relevant fall-out of light absorbing aerosols onto the sea ice increased by 17 ± 5 % during the past fifty years. A deposition of additional 3 ± 1 % of solar radiation in the melting region results that accounts for the ice retreat. Recalling the important role of the ice loss for the terrestrial climate,[3,5,9] the precipitation of air pollution in the Arctic seems to be an important factor for the global warming.

Introduction

The relationship between Arctic warming and sea ice loss is not well understood.[14] Some authors point out that is partly due to natural reasons.[2,3,15,16] There is evidence for climate changes in past centuries evolving on time scales of hundreds of years, accompanied by distinct changes of the Arctic sea ice [3,17-19] that may continue at the present. The loss of sea ice was also related to the North Atlantic Oscillation Index [20-22] and the Atlantic Driver [23,24]. On the other hand, evidence was reported for a large contribution of summertime atmospheric circulation of 60 % to the September sea ice loss.[13]. Notz and Stroeve reported that the observed Arctic sea ice loss directly follows anthropogenic $CO_2$ emissions, whereas climate model simulations for the sea ice retreat differ substantially.[25] The discussion presented below is based on the satellite observations of Ref. [11] for the daily sea ice extent and the data of Ref. [26] for the maximum and minimum sea ice area in recent decades.

For the analysis of the experimental information an analytic model with several simplifications is used:

- The data for the daily sea ice area are derived from the daily ice extent values, assuming that the ratio of the two quantities in each respective year is approximately constant.[26]
- Following Ref. [27] and averaging over seasonal changes, the ice thickness in the melting region is taken to be 1.5 m and approximately constant in the years 1979 to 2018. As a result the daily energy consumed by ice melting in the summer period is just proportional to the daily loss of sea ice area.
- For the melting energy of sea ice we only distinguish two sources: direct absorption of solar radiation of the ice surface and transport of heat to the melting zone by sea water and/or the atmosphere.
- The energy transport that obviously contributes to the ice melting in the summer period is estimated from the time evolution of the freezing energy ( = negative melting energy) of ice at the beginning and end of the winter period when solar radiation is negligible (for details see below).
- The solar radiation in the daily melting zone (i.e. difference of the southern edge of the melting region with maximum ice and the daily melting frontier) is estimated for a transmission of 0.49 of the atmosphere of the sun light and omitting geographical details of the Arctic region. In other words, the earth is treated as rotating sphere with known radius while the daily angular position of the sun is varying between the Tropic of Cancer (23.44° N) and Tropic of Capricorn (23.44° S). Geographical details of the ice are omitted. In terms of the model, the melting zone of the Arctic ice extends between the frontier line of the annual maximum of sea ice (i.e. minimum latitude at the end of the freezing period) and the daily melting frontier. The latter shifts to the north and reaches a maximum value at the end of the melting period (i.e. annual minimum of ice area). For example, for 1979 the latitude limits are 70.34 °N and 78.00 °N, respectively (71.37 °N and 81.32 °N in year 2017). Because of the various simplifications of the model the results reported below are of semi-quantitative nature, only.

Results and Discussion

The analysis was carried out for nine years in the period 1979 to 2018. An example for 1979 is presented in Figs. 1 and 2.

The reported data of Ref. [11] for the sea ice extent were smoothed to remove some of the experimental scatter; the data were converted to ice area values by the help of Ref. 26 for maximum and minimum ice area (dash-dotted black curve in Fig. 1), assuming that ice extent and ice area are approximately proportional [28]. The ice area is shown in the Fig. to rise from January 1$^{st}$ (day = 1) to a maximum at March 6$^{th}$ (day = 62, end of the freezing period). For subsequent days the area declines to a minimum at day = 255 (September 12$^{th}$, end of melting period). The daily loss of ice area is illustrated by the red curve in Fig. 1 that simply represents the daily change of the dash-dotted line. We note that maximum daily melting occurred in July 1979 on day 199, notably later than the beginning of summer (day 172, see red curve). Comparing the reported data set for 1979 and 2018 (data not shown) the decrease of the annual area minima (factor 0.51) is more pronounced than that of the annual maxima (factor 0.90); i.e. the seasonal variation of sea ice area in individual years is rising in the period 1979 to 2018 (factor 1.13).

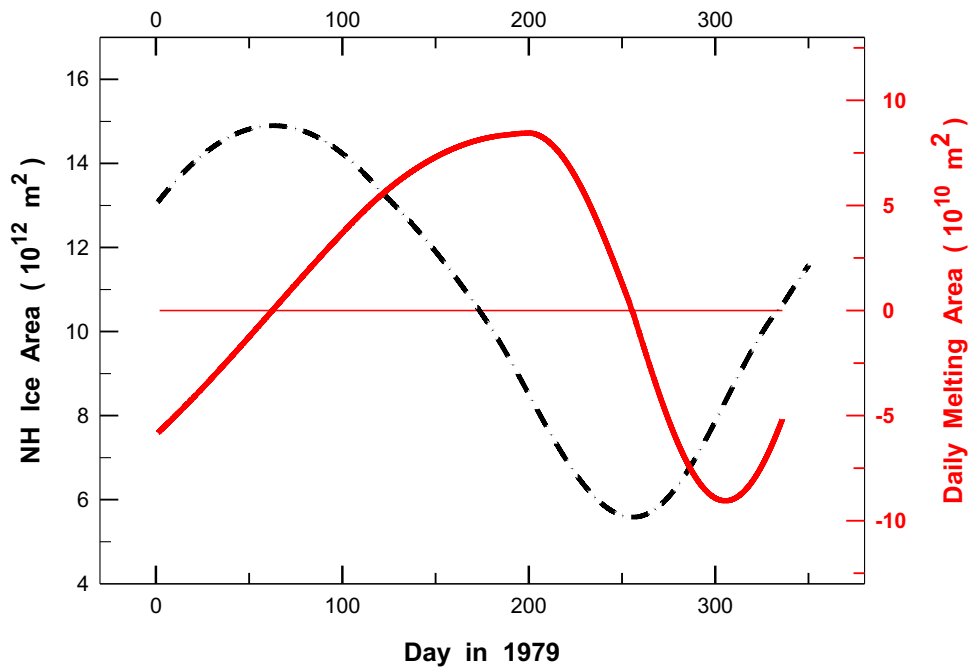

Fig. 1: Melting kinetics of Arctic sea ice in 1979 [11, 26]: Arctic sea ice area (black dash-dotted line, l.h.s. ordinate scale) and daily melting area (solid red curve, r.h.s ordinate scale); see text.

Calculated information on the energy flows in the melting zone of the Arctic in 1979 as derived from the ice kinetics is presented in Fig. 2. The energy consumed for melting is defined to be negative (= heat loss of the ice in the melting zone) in the summer period and illustrated by the solid red curve; positive values indicate freezing. The red curve is just proportional to the red curve in Fig. 1 (with inverted sign just by definition), because of the known latent heat of ice and the assumed constant thickness (seasonal variations of the ice thickness are omitted). Fig. 2 also shows the energy of solar radiation in the melting region (dashed green curve). Comparing the two curves we note that daily melting still increases after day 150 when sunshine in the region is already declining. Rising daily heat transport to the melting zone and / or increasing absorption of solar radiation by the ice surface is required to explain the delayed melting behaviour.

We also note in Fig. 2 positive values of the melting energy curve for the freezing of sea water in the winter period (see solid red line). The finding illustrates energy loss by heat transport, since the solar input is vanishing (or negligible) for these days. This negative heat transport is shown in the Fig. for days < 50 and days > 280 (solid blue curve). It just represents the corresponding parts of the red curve with inverted sign.

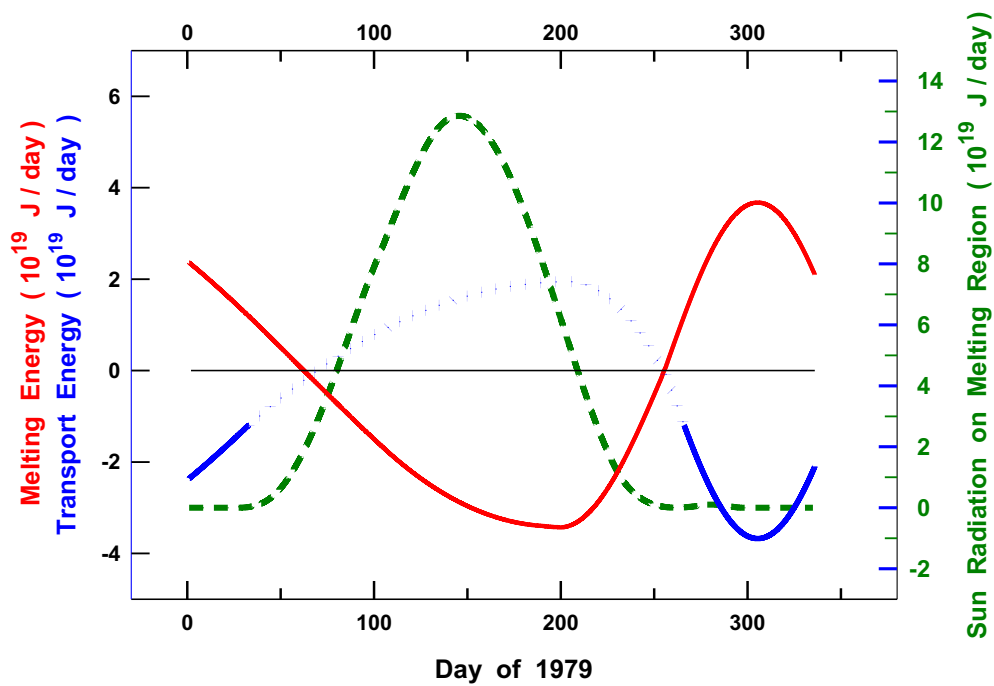

Fig. 2: Melting in 1979: daily melting energy (solid red curve, l.h.s. scale) and estimated mean daily heat transport (solid blue and dotted blue line, l.h.s. scale); available daily solar radiation for the melting region (short dashed green curve, r.h.s. scale) is shown for comparison; see text. For melting the sign of the melting energy is defined to be negative, while heat transport is chosen to be positive.

In the summer period the heat transport from the south contributes to melting, while solar input accounts for melting only in part. In other words, heat transport changes sign. The dotted part of the blue curve in Fig. 2 is an estimate of this energy transport in the summer period, i.e. the contribution of heat transport to the ice melting. It is a rough estimate, only, and obtained by interpolation. Two parabolas are fitted respectively to the left and right branch of the solid blue line in the figure, with the joint parabolic maximum taken to occur around day 2000 (close to the maximum daily melting). The latter assumption is sufficient to determine the maximum amplitude. The procedure suggests a maximum daily heat transport of approximately $(2 \pm 0.7) \times 10^{19}$ Joule/day in 1979. The maximum daily melting, on the other hand, requires $3.4 \times 10^{19}$ Joule/day (compare red line in Fig. 2). The difference of the two numbers is noteworthy. Quite obviously direct input of solar radiation to the sea ice is required to explain the reported melting kinetics; i.e. absorption of sun light by the sea ice surface.

Similar data are deduced from the reported ice data for subsequent years (Ref. 11, supported by Ref. 26). An overview is presented in Figs 3 and 4 showing annual summaries over the respective melting periods of nine years (day ~ 64 to day ~ 255). The required annual melting energy is depicted in Fig. 3 (open red circles and solid red line) together with the estimated annual contribution of energy transport for the melting (blue triangles and blue dotted curve).

The lines are linear fits to the data for nine years in 1979 to 2018. The annual energy demand for melting of sea ice (absolute value) is found to rise from $3.8 \times 10^{21}$ J to $4.3 \times 10^{21}$ J (compare fitted red line). The rise by $13 \pm 1$ % (red curve) is directly related to the increasing melting amounts of the sea ice, corresponding to the difference of the annual maxima and minima of sea ice area (and volume). The annual minima decrease more rapidly than the maxima (not shown in the figure) so that the difference is growing. The available solar energy for melting in the respective years is calculated to rise from $1.3 \times 10^{22}$ J to $1.6 \times 10^{22}$ J (increase of approximately 20%, see dashed green line). The positive slope may be compared with the growing annual ice retreat (1979: $9.3 \times 10^{12}$ m²; 2018: $10.5 \times 10^{12}$ m²). The required annual contribution of heat transport appears to be approximately constant with $(2.2 \pm 0.7) \times 10^{21}$ J (see dotted blue line in Fig. 3). The latter finding may be compared with two opposing factors: the rising global warming on the one hand and the north shift of the melting region on the other one, hindering heat transport from the south.

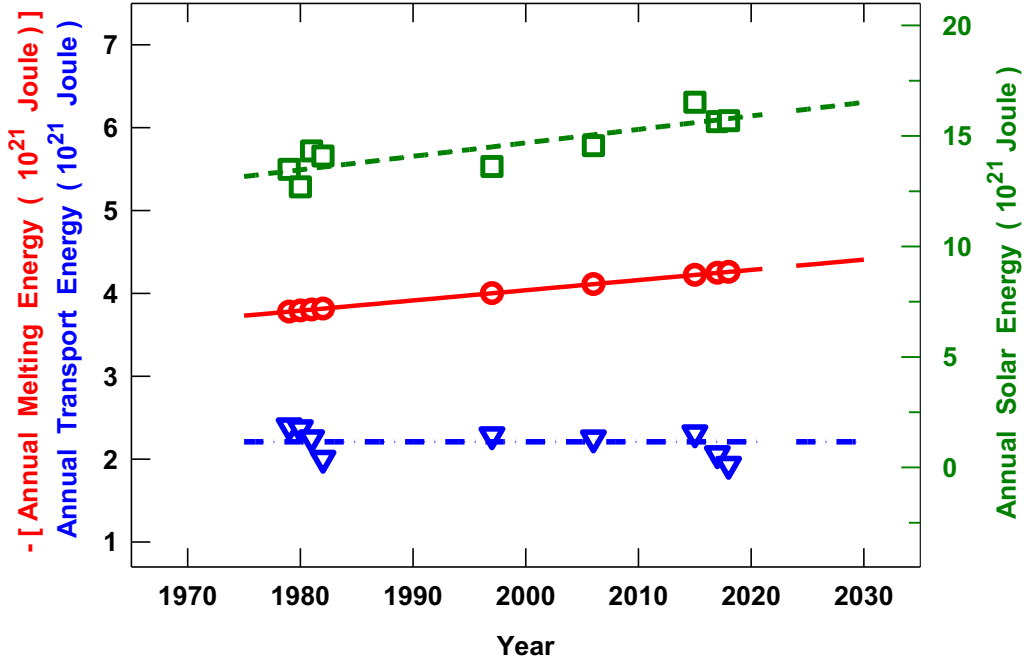

Fig. 3: Annual energy amounts in the melting periods of Arctic sea ice: melting energy (absolute value; open red circles and fitted solid red line) and estimated transport energy (blue triangles and fitted blue dash-dotted line) in the respective melting periods. The available energy of sunshine in the melting region during the melting period of the respective years is also shown (green broken line; r.h.s. ordinate scale); see text.

Air temperatures in the pole region are reported to rise much more than the global mean temperature (factor of 2 or more) [29] while the sea water directly beneath the permanent polar ice shield (i.e. water-ice interface) is obviously constant at ~272 K.

An increasing absorption of sun light by the ice surface may be inferred from the data of Fig. 3. This conclusion is illustrated in more detail by Fig. 4. The difference of annual melting energy and annual energy transport as derived from our model is plotted relative to available solar radiation in the melting region (open red circles, broken red line). The data suggest that an increase of ~3 % of solar energy is required for the ice retreat in the years 1979 to 2018, i.e. growing surface absorption. The ratio of absorbed solar energy and sea ice area, i.e. specific surface absorption is also shown (full blue triangles, dash-dotted blue line). The 17% rise of the curve (170 MJ/m$^2$ and 200 MJ/m$^2$ in 1979 and 2018, respectively) is directly related to the growing ice melting.

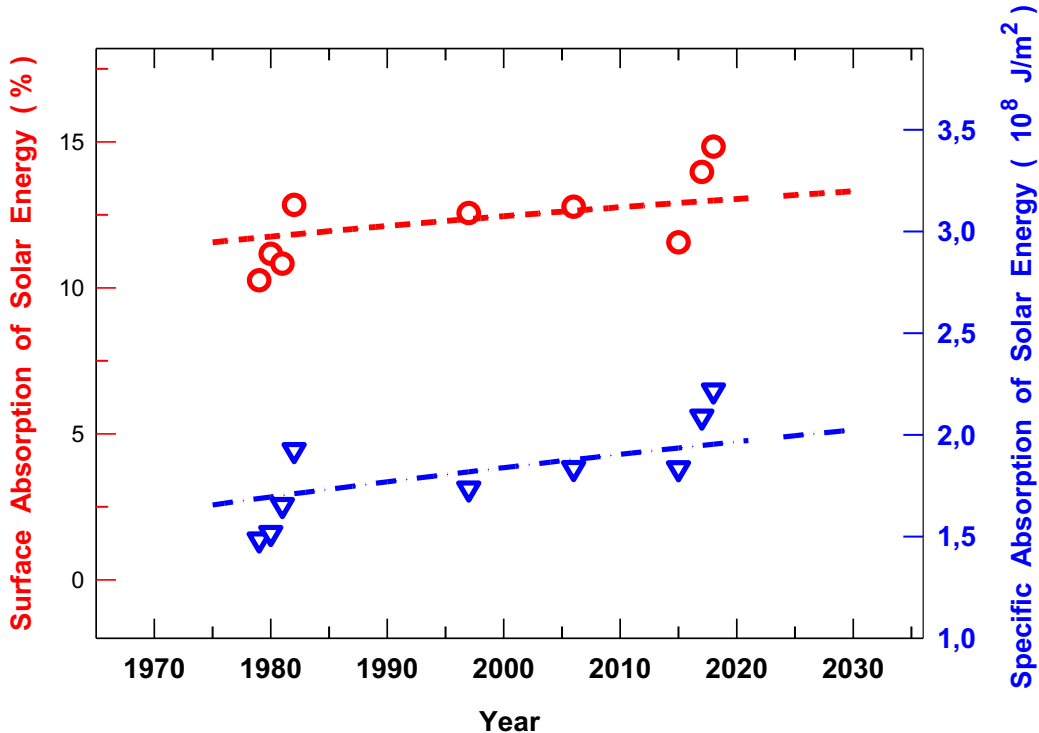

Fig. 4: Average annual surface absorption of solar radiation in the melting region relative to incident radiation (broken red line and open red circles, l.h.s ordinate scale); specific absorption of the solar input, i.e. annual mean ratio of absorbed solar energy and melting area of the sea ice (dash-dotted blue line and blue triangles, r.h.s. ordinate scale) in 1979 to 2018.

We recall that our data analysis refers to a constant value of the average ice thickness in the annual melting periods. Inspection of the reported ice extent data shows that the beginning of the freezing period (day ~255 in years 1979 to 2018) is approximately constant while the duration of the freezing period seems to increase slightly by 3 days. The finding supports the finding that the average ice thickness of the melting zone was not shrinking in recent decades (seasonal changes omitted).[27]

## Conclusion

Using a simple model for the retreat of Arctic sea ice by the analysis of reported ice data we derive evidence for a significant role of surface pollution. The latter obviously originates from the atmosphere and gives rise to an increase of ~3 % of solar energy in the melting region. The finding leads to several important questions:

(i) How large is the contribution of the Arctic sea ice retreat to the global warming?

(ii) To what extent is the sea ice retreat caused by the concentration rise of greenhouse gases (GHG)?

(iii) What is the relationship between the Arctic temperature anomaly, i.e. rise of the surface temperature of the Arctic (exceeding the global warming by a factor of two or more) and the melting of Arctic ice?

(iv) What kind of atmospheric pollution is relevant for the retreat of Arctic sea ice?

Although it is generally accepted that the ice retreat contributes to the global warming via an albedo decrease [7,29] there is disagreement how large this effect is.[30] The present authors have estimated that roughly 50 % of the temperature rise of ~1.2 °C of the northern hemisphere is due to the loss of Arctic ice since 1979.[5,6] Consistently we calculate that the GHG increase contributes only ~25 %. Feedback of global warming and the growth of GHG in the atmosphere, therefore, is considered to be of minor importance for the polar ice melting, since the Arctic temperature increase of ~6 % with respect to the freezing point of ice may not be a crucial effect. Consistently, the Arctic temperature rise appears to be dominantly caused by the ice retreat [3,31] and not vice versa.

Although different air pollutants contribute to the warming of the atmosphere [29,32] or cooling of the surface, several authors pointed to the important role of black carbon (from fuel combustion and biomass burning) and mineral dust for the darkening of the northern snow and ice surface.[29,32,33] In the light of these arguments it is suggested that light absorbing aerosols in the atmosphere should be considerably reduced to limit further global warming .